# Photodesorption of water ice from dust grains and thermal desorption of cometary ices studied by the INSIDE experiment

Alexey Potapov[1], Cornelia Jäger[1], and Thomas Henning[2]

[1]*Laboratory Astrophysics Group of the Max Planck Institute for Astronomy at the Friedrich Schiller University Jena, Institute of Solid State Physics, Helmholtzweg 3, 07743 Jena, Germany, email: alexey.potapov@uni-jena.de*
[2]*Max Planck Institute for Astronomy, Königstuhl 17, D-69117 Heidelberg, Germany*

A new experimental set-up INterStellar Ice-Dust Experiment (INSIDE), was designed for studying cosmic grain analogues represented by ice-coated carbon- and silicate-based dust grains. In the new instrument, we can simulate physical and chemical conditions prevailing in interstellar and circumstellar environments. The set-up combines ultrahigh vacuum and low temperature conditions with infrared spectroscopy and mass spectrometry. Using INSIDE, we plan to investigate physical and chemical processes, such as adsorption, desorption, and formation of molecules, on the surface of dust/ice samples. First experiments on the photodesorption of water ice molecules from the surface of silicate and carbon grains by UV photons revealed a strong influence of the surface properties on the desorption yield, in particular in the monolayer regime. In the second experiment, the thermal desorption of cometary ice analogues composed of six molecular components was studied for the first time. Co-desorption of $CO_2$ and $CH_3OH$ with $O_2$ indicates that at high $O_2$ concentrations in cometary or interstellar ices "heavy" ice molecules can be partly trapped in $O_2$ and release to the gas phase much earlier than expected. This effect could explain astronomical detections of complex organic molecules in cold dense interstellar clouds.

## 1. Introduction

The interstellar medium (ISM) attracts our great attention as the place where stars and planets are born and from where, possibly, the precursors of life have come to Earth. The ISM is filled with a tenuous amount of materials, such as gas and dust. The dust, about 1% by total mass, consists of carbon- and silicate-based compounds (Dorschner & Henning 1995; Henning & Salama 1998; Draine 2003,2009; Henning 2010). In cold and dense environments of the ISM, such as dense molecular clouds and outer parts of circumstellar envelopes and disks, dust grains are mixed with and covered by molecular ices consisting mainly of a number of simple molecules: $H_2O$ (the major compound), CO, $CO_2$, $NH_3$, $CH_3OH$, $CH_4$, $O_2$ and some others including a small amount of complex molecules, and polycyclic aromatic hydrocarbons (for reviews see (Allamandola et al. 1999; van Dishoeck 2014)). It is generally accepted that large bodies in circumstellar disks, such as planets, planetesimals, asteroids, and comets, are formed by aggregation of dust grains. The presence of an ice layer, particularly of polar ices, increases the efficiency of the grain aggregation (Boogert, Gerakines, & Whittet 2015; Min et al. 2016). The cometary ice composition is thought to be similar to that of the interstellar ice mantles (Mumma & Charnley 2011).

Chemical processes in the ISM can be divided into two groups, gas phase reactions and reactions on the surface of interstellar grains. These two reaction routes interact through adsorption/desorption processes. Physical and chemical processes in/on grains are triggered by UV-irradiation, cosmic rays, thermal processing, atom addition, and reactions with radicals and lead to the formation of new molecules and, possibly, their release to the gas phase. The schematic Figure 1 shows cosmic grains mixed with molecular ices and the main sources of their processing in astrophysical environments. The influence of the structure, temperature, and composition of grains on the physical and chemical processes on their surfaces is of great importance for the general understanding of astrochemical networks, the creation of planetary systems, and the appearance of life on Earth.

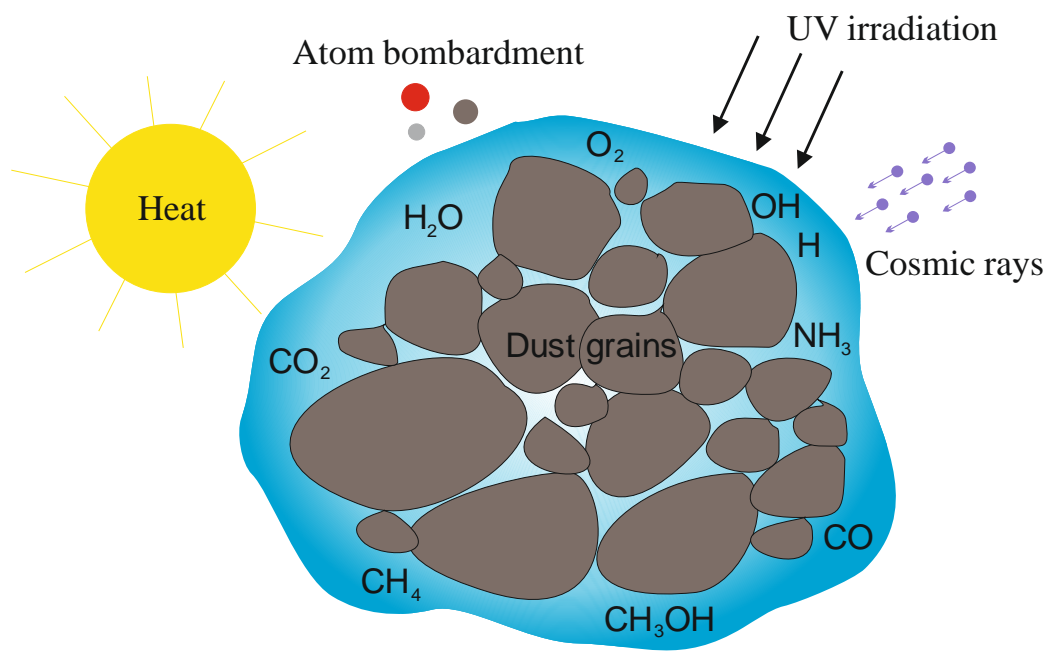

Figure 1. Schematic figure showing cosmic grains mixed with molecular ices and the main sources of their processing in astrophysical environments.

Outside the sublimation distances, the dominant destructive process for the ice coverage of grains in circumstellar disks is supposed to be UV photodesorption. It was shown that photodesorption efficiently destroys water ice in optically thin discs (Grigorieva et al. 2007). Photodesorption yields have been measured for a number of ices, such as $H_2O$ (Westley et al. 1995; Öberg et al. 2009a; Cruz-Diaz et al. 2018), CO (Öberg et al. 2007; Öberg, van Dishoeck, & Linnartz 2009b; Muñoz Caro et al. 2010; Chen et al. 2014), $CO_2$ (Öberg, et al. 2009b; Yuan & Yates 2013), $N_2$ (Öberg, et al. 2009b; Fayolle et al. 2013), $O_2$ (Fayolle, et al. 2013; Zhen & Linnartz 2014), and $CH_3OH$ (Bertin et al. 2016). In these studies, ices were deposited on standard laboratory substrates, such as gold or KBr, that are not comparable to the dust surfaces present in the ISM.

To the best of our knowledge, the only study on the photodesorption of a monolayer of water ice coating an amorphous carbon foil by 193 nm photons was presented by Mitchell et al. (Mitchell et al. 2013). From the results presented there, one can conclude that water ice photodesorbs more efficiently from a carbon surface than from a standard substrate. What can also be mentioned in this context are the studies of the UV-induced formation of CO and $CO_2$ molecules on carbon grains (Mennella et al. 2006; Fulvio, Raut, & Baragiola 2012; Shi, Grieves, & Orlando 2015) and the study of the effect of an amorphous water-rich magnesium silicate on the chemical evolution of UV-irradiated methanol ices (Ciaravella et al. 2018).

There is a huge number of laboratory studies of temperature-programmed desorption (TPD) of interstellar ice analogues (for reviews see (Burke & Brown 2010; Theule et al. 2013)), involving desorption from silicate and carbon surfaces (Ulbricht et al. 2006; Thrower et al. 2009; Noble et al. 2012a; Noble et al. 2012b; Clemens et al. 2013; Smith et al. 2014; Doronin et al. 2015; Suhasaria, Thrower, & Zacharias 2015; Smith, May, & Kay 2016; Suhasaria, Thrower, & Zacharias 2017; Potapov, Jäger, & Henning 2018a). The majority of these studies are devoted to the TPD of pure ices or binary mixtures, however, interstellar and circumstellar ices are more complex. Recently, the results of TPD experiments of cometary ice analogues composed of five molecular components including $H_2O$, CO, $CO_2$, $CH_3OH$, and $NH_3$, the main constituents of interstellar and circumstellar ices, were presented (Martin-Domenech et al. 2014). The Rosetta mission to the comet 67P/Churyumov–Gerasimenko has shown that the abundance of molecular oxygen in the coma of the comet is much higher than expected and ranges from one to ten percent relative to $H_2O$ with a mean value of 3.8 percent. The oxygen was probably incorporated into the nucleus during the comet's formation (Bieler et al. 2015). The local number densities in the coma of 67P vary spatially and temporally for different compounds (Hassig et al. 2015; Luspay-Kuti et al. 2015) and their relations to the comet structure and composition are not yet clear. The presence of $O_2$ in cometary and interstellar ices can alter the spectral properties of other ice components (Ehrenfreund et al. 1997; Muller et al. 2018) and their binding energies as it was demonstrated for $O_2$:CO ice mixtures (Acharyya et al. 2007). Thus, the presence of $O_2$ in ice can have an influence on the amount of species released to the gas phase in interstellar and circumstellar environments.

The paper describes a new ultra-high vacuum set-up, where carbon and silicate dust grains covered by molecular ices of variable compositions can be formed at different temperatures (down to 15 K) and processed by UV irradiation and thermal heat. Typical temperatures in dense molecular clouds are in the range of 10 – 50 K. Typical densities in such environments are $10^3$ to $10^7$ molecules cm$^{-3}$ corresponding to pressures between $10^{-14}$ and $10^{-10}$ mbar. Thus, the pressure of a few $10^{-11}$ mbar allows us to reproduce the physical conditions in dense interstellar clouds and to perform clean experiments without additional adsorption of species from the chamber volume. The set-up is movable and can be coupled to different irradiation sources including X-rays or energetic particles at synchrotron or accelerator facilities. Dust/ice samples can be studied in-situ by using Fourier transform infrared (FTIR) spectroscopy and mass spectrometry (MS).

A detailed technical description of INSIDE is presented in the sections 2 and 3. First results on the UV photodesorption of water ice from the surface of silicate and carbon grains and on the temperature-programmed desorption of cometary ice analogues composed of six molecular components including $H_2O$, CO, $CO_2$, $CH_3OH$, $NH_3$, and $O_2$ from a KBr substrate are presented in sections 4 and 5, correspondingly.

## 2. Technical description of INSIDE

The set-up shown in Figures 2 and 3 has a vertical configuration and consists of two chambers separated by a gate valve. The upper ultrahigh vacuum (UHV) chamber was designed for measurements of interstellar dust/ice analogues at ultrahigh vacuum ($4\times10^{-11}$ mbar at room temperature) and low temperature (down to 15 K) conditions. The lower high vacuum chamber (pressure $\approx 10^{-8}$ mbar) is used for the annealing and extraction of dust samples without breaking the UHV conditions in the main chamber.

The low pressure in the UHV chamber is attained by pumping the chamber with a combination of a turbomolecular pump with a pumping capacity of 300 l/s (Oerlikon) backed up by a scroll pump and a NEXTorr pump (SAES) with a pumping capacity of 500 l/s for $H_2$ combining a non-evaporable-getter element with an ion pump element. The chamber has to be baked at 100°C for two weeks. Baking at higher temperatures is not possible due to the temperature limit of the coldhead. The cooling down of the sample to 15 K is achieved by mounting the sample holder on a coldhead connected to a closed-cycle helium cryostat (Advanced Research Systems).

The home-made sample holder with a KBr substrate covered by a layer of dust grains produced in another experiment (see section 3) is moved into the lower chamber from the bottom by using a magnetically coupled rotatable translator. In order to get rid of possible adsorbate layer, the sample is typically annealed at 200°C for two hours, using a home-made annealing setup. After that, the gate valve is opened and the sample holder is moved further up and screwed on the coldhead. The translator can be removed and after closing the gate valve, the original pressure of $10^{-11}$ mbar is reached in the chamber after about an hour.

The temperature is measured by a temperature diode permanently fixed on the coldhead. An additional diode was fixed on the sample holder near the sample and the coldhead diode was calibrated with respect to it. The second diode was removed after the calibration to allow the translation of the sample holder during the sample change procedure. All temperatures mentioned in the paper are the temperatures on the sample holder.

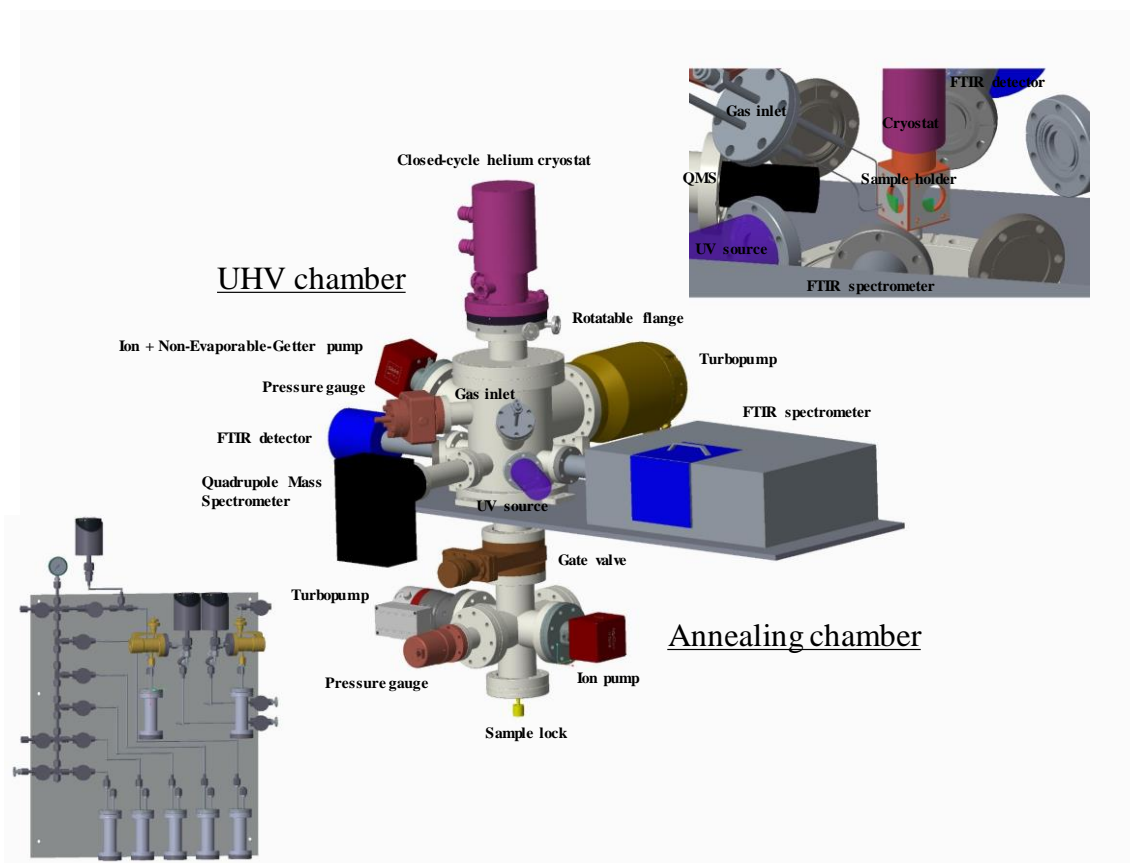

Figure 2. Schematic of the INSIDE experimental set-up. Inset top-right: UHV chamber, internal view. Inset down-left: gas inlet system.

When the holder with the grain sample is fixed at the coldhead, it can be cooled down to 15 K. The coldhead is mounted onto the UHV chamber by a rotatable flange that allows us to turn the sample in different working positions. The flange in individually evacuated by a small serial turbo pump (HiPace 10, Pfeiffer) backed up by a membrane pump.

The gas inlet system of INSIDE is presented in the inset of Figure 2. It has two separate lines. The first one allows the mixing of up to five different gases. The second one is equipped with special gold seals in the leakage valves and is used for corrosive species, such as ammonia.

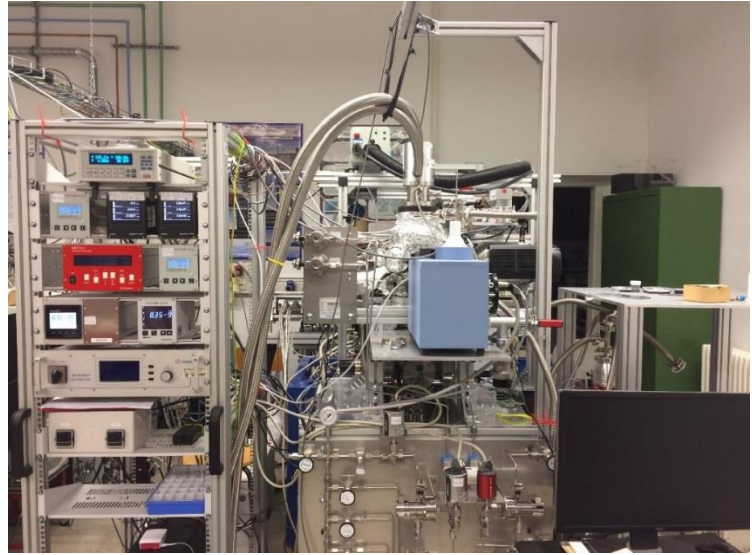

Figure 3. A photo of the developed set-up INSIDE. The central part shows the UHV chamber equipped with the cryostat, the small chamber containing the external IR (MCT) detector, and the gas inlet system with the two different gas lines. The rack at the left side contains the temperature controller of the cryostat, turbo pump controllers, and monitors showing the pressure in various chambers.

A schematic view of the UHV chamber is shown in Figure 4. In the present configuration of INSIDE, the following procedures can be performed:

1) Production of dust/ice samples

Gas mixtures can be deposited onto grains at different temperatures through two 0.5 mm diameter capillary tubes displayed in the inlet of Figure 2. Two leakage valves are connected to the two different gas lines. The deposition through both lines can be done simultaneously.

2) UV processing

The dust/ice sample can be irradiated by UV photons. Two UV lamps separated from the main chamber by a $MgF_2$ window can be used: a home-made open hydrogen flow discharge lamp with a flux of $10^{13}$ photons s$^{-1}$ cm$^{-2}$ emitting at Ly$\alpha$ at 121 nm (10.2 eV) and a broadband deuterium lamp (L11798, Hamamatsu) with a flux of $10^{15}$ photons s$^{-1}$ cm$^{-2}$. The lamp has a broad spectrum from 400 to 118 nm with the main peaks at 160 nm (7.7 eV) and at about 122 nm (10.2 eV) (with the intensity of about 20% of the 160 nm peak). The signal above 170 nm is weak (about 10% of the total intensity). The main peaks correspond to the emission of molecular and atomic hydrogen. Such an emission spectrum can be a good analog of UV fields in interstellar clouds but also a broad UV spectrum is useful for modelling photophysical and photochemical processes in protostellar envelopes and planet-forming disks, where UV photons come from a central star having a continuous spectrum. UV fluxes of the lamps were measured in a separate experiment by the method based on the photoeffect (Fulvio et al. 2014). The uncertainty of the UV fluxes is about 30%.

3) Thermal annealing

The dust/ice sample can be heated up to 300 K with a controlled heating ramp rate.

4) Analytical characterisation

a) An FTIR spectrometer (Vertex 80v, Bruker) is used to perform in-situ spectroscopy in transmission mode. The spectrometer is equipped with a MCT-detector working in the spectral range of $12000 - 600$ cm$^{-1}$. The FTIR line is separated from the main chamber by two ZnSe windows and pumped by a dry backing pump.

b) Species released from the sample surface can be analysed in the gas phase by a quadrupole mass spectrometer (QMS) equipped with a Faraday and a multiplier detector (multiplication factor of $10^3$) having a mass range up to 300 amu and a partial pressure sensitivity of $5\times10^{-14}$ Torr (HXT300M, Hositrad).

c) An UV-VIS spectrometer can be coupled to the chamber, thus obtaining a third detection technique and extending the range for spectroscopic measurements.

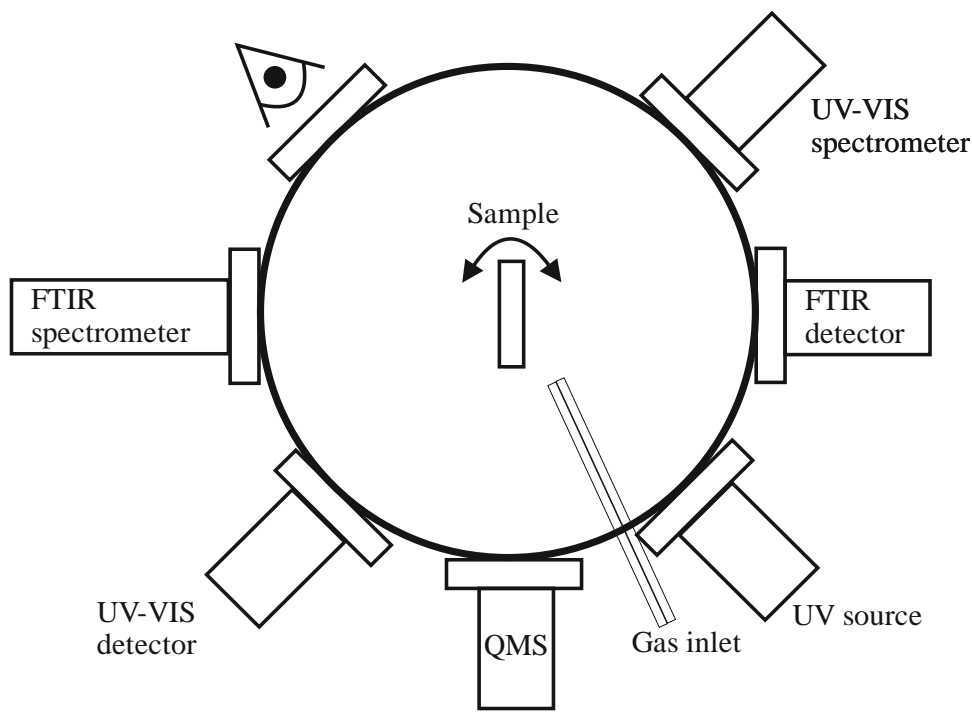

Figure 4. Schematic representation of the main chamber of INSIDE.

## 3. Production of dust grain analogues

Dust grain analogues are produced in a laser ablation set-up, which is able to condense solid carbonaceous and siliceous grains and gases at temperatures between 300 K and 8 K. The setup is detailed elsewhere (Jäger et al. 2008; Jäger et al. 2009; Potapov, et al. 2018a; Potapov et al. 2018b). The gas-phase deposition of nanometre-sized amorphous grains is achieved by pulsed laser ablation of graphite or metal targets composed of Mg, Fe, and Si and subsequent condensation of the evaporated species in a quenching atmosphere of a few mbar He and $H_2$ for amorphous carbon grains or He and $O_2$ for amorphous silicate grains. The condensed grains are extracted adiabatically from the ablation chamber through a nozzle into a second low pressure chamber ($10^{-3}$ mbar) to decouple the grains from the gas. A second extraction through a skimmer into a third chamber ($10^{-6}$ mbar) generated a particle beam that is directed into a fourth, cryogenic, deposition chamber ($10^{-6}$ mbar), where the grains are deposited onto a substrate. The thickness of the grain deposit is controlled by a microbalance. The composition of the employed silicates was determined by HRTEM and EDX studies.

The grain structure can be described as amorphous. The individual particles are very small (a few nm) and the largest particle agglomerates are in the range of a few tens of nm. The morphology of dust particles on a substrate can be understood as a porous layer of rather fractal agglomerates. The porosity of the layer can be as high as 90% (Sabri et al. 2015). The surface of the dust is very large, but the area cannot be measured exactly. Figure 5 shows a SEM (scanning electron microscopy) image of silicate grains produced in our set-up.

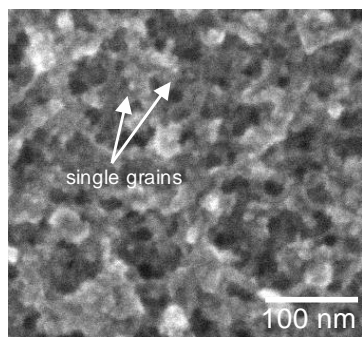

Figure 5. A SEM image of the porous structure of silicate grains produced by laser ablation. The individual particles are very small (a few nm) and can hardly be resolved from the image. Carbonaceous grains produced by laser ablation show a very similar morphology.

It is worth to note that carbon grains are composed of small, strongly bent graphene layers or fullerene fragments, which are linked by aliphatic bridges. Fullerene molecules of different sizes and shapes can also be contained (Jäger, et al. 2008; Jäger, et al. 2009). Fullerene-like grains, as they are typically considered in the literature, are characterized by refractory and disordered carbonaceous structure. The bent graphene layers and aliphatic bridges can have a high number of defects and dangling bonds.

### 4. Photodesorption of water ice from the surface of dust grains

Figure 6 shows the IR transmission spectra of $MgFeSiO_4$ and $^{13}C$ grains deposited on KBr substrates, the samples used in the present study. The spectra were taken in the UHV chamber of INSIDE after their exposure to air and annealing in the lower chamber of INSIDE. In the silicate spectra, one can see the band related to the Si–O stretching vibration at 990 $cm^{-1}$ and additional bands due to interaction with air: around 3500 $cm^{-1}$ corresponding to O–H stretching in associated Si–OH groups and around 1500 $cm^{-1}$ corresponding to magnesium carbonate. In the carbon spectra the assignable bands are due to C-H stretching vibrations corresponding to the saturated −$CH_3$ and −$CH_2$ groups, C=O and C=C stretching, and C-H deformation.

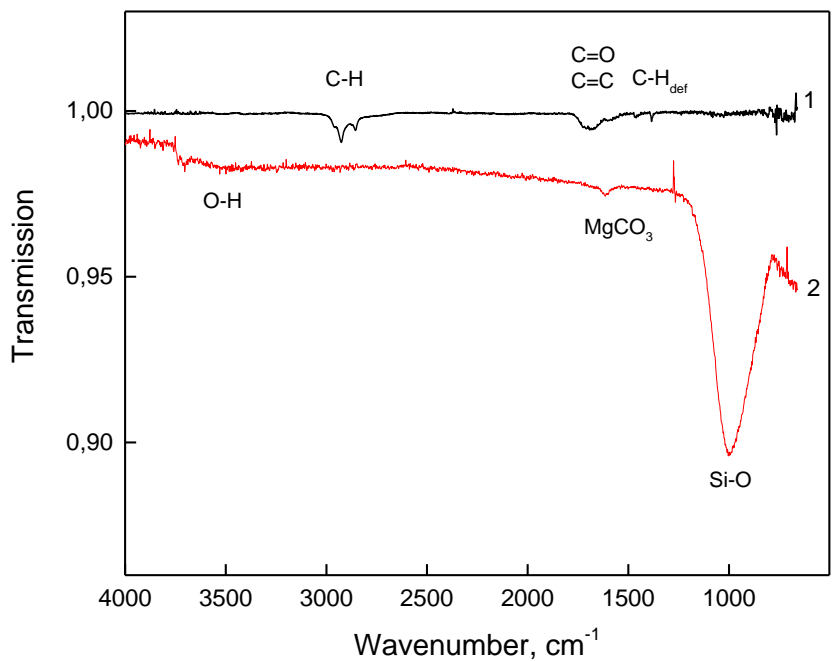

Figure 6. IR transmission spectra in the UHV chamber: 1 - carbon grains (70 nm thickness), 2 - MgFeSiO$_4$ grains (150 nm thickness). Spectrum 2 is offset for visibility.

For these experiments, 3 and 66 monolayers (ML) of water ice were deposited onto the surface of amorphous MgFeSiO$_4$ and amorphous $^{13}$C grains at 15 K. The deposition rate was about 10 ML min$^{-1}$. The thickness of the water layer was determined from the 3,07 μm water band area using the band strength of 2x10$^{-16}$ cm/molecule (Hudgins et al. 1993). After deposition, the samples were irradiated at 45 degrees by the deuterium lamp for two hours with the final UV fluence of 5x10$^{18}$ photons cm$^{-2}$. Figure 7 shows IR transmission spectra before and after the UV irradiation of H$_2$O ice deposited on carbon grains. Figure 8 presents the dependence of the ice thickness determined from the IR spectra on the UV fluence. The photodesorption yields were determined from linear fits of these dependencies for the samples studied and they are given in Table 1. The total experimental uncertainty estimated from the uncertainty of the photon flux, linear fit, and repeated experiments is ∼40% for the total photodesorption rate.

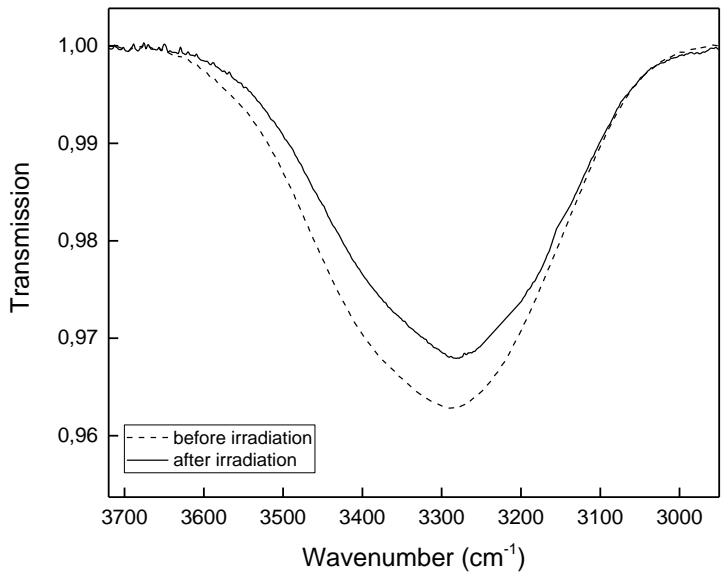

Figure 7. IR transmission spectra before and after the UV irradiation of $H_2O$ ice (66 ML sample) deposited on carbon grains at 15 K.

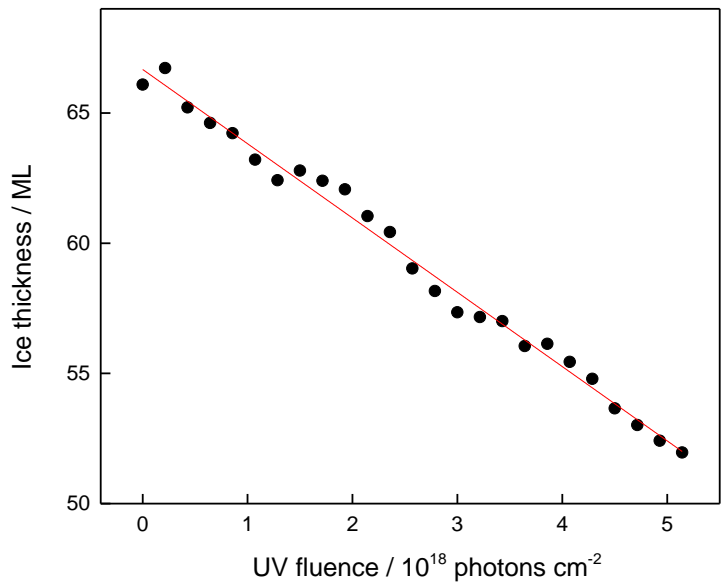

Figure 8. UV fluence dependence of the thickness of $H_2O$ ice (66 ML sample) deposited on carbon grains at 15 K and its fit by a linear function.

Table 1. Photodesorption yields of H$_2$O measured in this study.

| Thickness (ML) | Surface | Photodesorption yield (10$^{-3}$ molecule/photon) |
|---|---|---|
| 66 | MgFeSiO$_4$ grains | 1.7 |
|    | $^{13}$C grains | 2.8 |
| 3  | MgFeSiO$_4$ grains | 0.5 |
|    | $^{13}$C grains | 4.8 |

An empirical dependence of the photodesorption yield on the temperature ($T$) and ice thickness ($x$) $Y_{pd}(T, x) = 10^{-3}(1.3 + 0.032 \times T)(1 - e^{-x/l(T)})$, where $l$ is a temperature-dependent ice diffusion parameter, was determined previously (Öberg, et al. 2009a). Using this dependence, we obtained the values for the photodesorption yield of H$_2$O at 15 K of ~1.8×10$^{-3}$ and 1.7×10$^{-3}$ molecule/photon for 66 and 3 ML correspondingly. Thus, the experimental values of the photodesorption yields for 66 ML of H$_2$O on MgFeSiO$_4$ grains is in agreement with the literature data. The binding energies of H$_2$O molecules on amorphous silicates and standard substrates, such as CsI and gold, experimentally determined by using the TPD technique were found to be quite similar varying within a value of about 15% (Penteado, Walsh, & Cuppen 2017). Thus, our result indicates that water ice is weakly bound on the surface of silicates. The much lower yield for 3 ML can be explained by a very large surface of grains that can be one to two orders of magnitude larger than the nominal surface of the substrate, on which grains are deposited (Potapov, et al. 2018a). Therefore, we probably have a sub-monolayer ice coverage of the dust instead of "nominal" 3 ML. Using the thickness of 0.3 ML instead of 3 ML in the empirical equation above, we obtain a photodesorption yield of 0.5 in agreement with our experimental value.

The photodesorption yield for H$_2$O on carbon grains is noticeably higher compared to silicate grains. For 66 ML, it is still in the limits of the total experimental uncertainty. However, for the 3 ML it is 9 times higher as compared to MgFeSiO$_4$ grains and 3 times higher as compared to the literature data. A reason for this finding could be the differences in the surface properties of carbon and silicate grains. The fullerene-like carbon grains used in the present study are typically considered to be hydrophobic, whereas silicates are known to have hydrophilic and hydrophobic surface groups. The finding of stronger bound water molecules to silicate grains in comparison to fullerene-like carbon grains was demonstrated in our recent study (Potapov, et al. 2018a). However, this result needs further investigations.

In the experiments with carbon grains, the formation of $^{13}$CO$_2$ was observed (see Figure 9). This result confirms that the dust surface is active. The formation of CO and CO$_2$ by UV irradiation of water ice covering the surface of carbon grains or layers is already well-known (Mennella, et al. 2006; Fulvio, et al. 2012; Shi, et al. 2015). It was also demonstrated that other energetic processes, such as ion and proton bombardment, lead to the same result (Mennella, Palumbo, & Baratta 2004; Raut et al. 2012; Sabri, et al. 2015). The addition of O/H atoms onto the surface of bare carbon grains makes one more step toward the molecular complexity leading to the formation of $H_2CO$ (Potapov et al. 2017). The formation cross section of $CO_2$ obtained in our study following the procedure described in (Mennella, et al. 2006) is $0.3 \pm 0.04 \times 10^{-18}$ cm$^{-2}$ and is in good agreement with the values provided there.

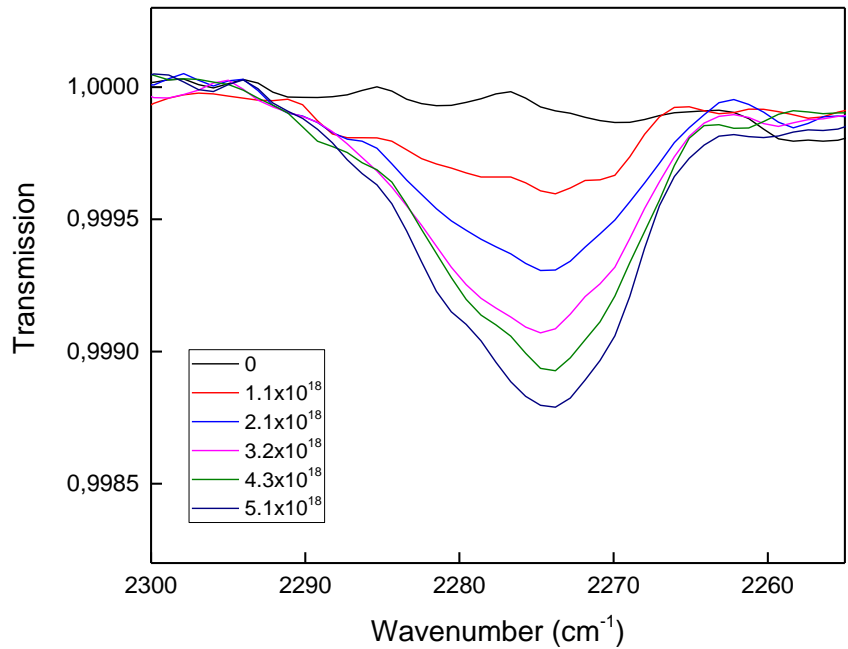

Figure 9. The $^{13}CO_2$ stretching mode of $H_2O$ ice deposited on carbon grains as a function of the UV fluence.

5. **TPD of pre-cometary ice composed of six components**

For these experiments, ices composed of five ($H_2O$, CO, $CO_2$, $CH_3OH$, $NH_3$) and six ($H_2O$, CO, $CO_2$, $CH_3OH$, $NH_3$, and $O_2$) molecules were deposited at 15 K onto a KBr substrate. $H_2O$, CO, $CO_2$, $CH_3OH$, and $O_2$ were premixed and $NH_3$ was deposited simultaneously with the mixtures through the separate line of the gas inlet system. We tried to reproduce the mixing ratios (by number of molecules) observed for the comet 67P/Churyumov–Gerasimenko (Altwegg et al. 2017; KathrinAltwegg private communication). There is a remarkable diversity in the mixing ratios of molecular ices in comets. Abundances for each species with respect to $H_2O$ range between one and two orders of magnitude (Dello Russo et al. 2016). We chose the comet 67P as the most recent object. The mixing ratios for 67P and used in our experiments are presented in Table 2. Figure 10 shows a spectrum of a mixture of six ices. The vibrational bands of $H_2O$, $CO$, $CO_2$, $CH_3OH$, and $NH_3$ are clearly observable and listed in Table 3. The ice thicknesses were determined from the IR band areas using the band strengths published elsewhere (Hudgins, et al. 1993). Due to a very weak oscillator strength of the $O_2$ vibration in the $H_2O$ matrix ($O_2$ itself has no characteristic infrared transitions), the corresponding band at 1559 cm$^{-1}$ (Ehrenfreund et al. 1992) was not observed. Therefore, mixing ratio of $O_2$ was estimated only from the $O_2$ pressure in the gas mixture before the deposition.

Table 2. Mixing ratios derived for the comet 67P (KathrinAltwegg private communication) and used in our experiments (by number of molecules, % relative to $H_2O$) and ice thicknesses.

|  | $H_2O$ | CO | $CO_2$ | $NH_3$ | $CH_3OH$ | $O_2$ |
|---|---|---|---|---|---|---|
| 67P ratios | 100 | 1.1 | 6 | 0.6 | 0.4 | 1 - 10 |
| Exp1 ratios | 100 | 1.5 | 7.7 | 0.4 | 0.6 | 0 |
| Thickness, nm | 230 | 3.4 | 17.7 | 1.0 | 1.4 |  |
| Exp2 ratios | 100 | 1.5 | 8.0 | 0.5 | 0.5 | 1 |
| Thickness, nm | 225 | 3.5 | 18 | 1.2 | 1.2 |  |
| Exp3 ratios | 100 | 1.5 | 6.9 | 0.7 | 0.5 | 4 |
| Thickness, nm | 226 | 3.5 | 15.6 | 1.6 | 1.2 |  |
| Exp4 ratios | 100 | 1.6 | 8.0 | 0.6 | 0.7 | 10 |
| Thickness, nm | 206 | 3.3 | 16.4 | 1.2 | 1.5 |  |

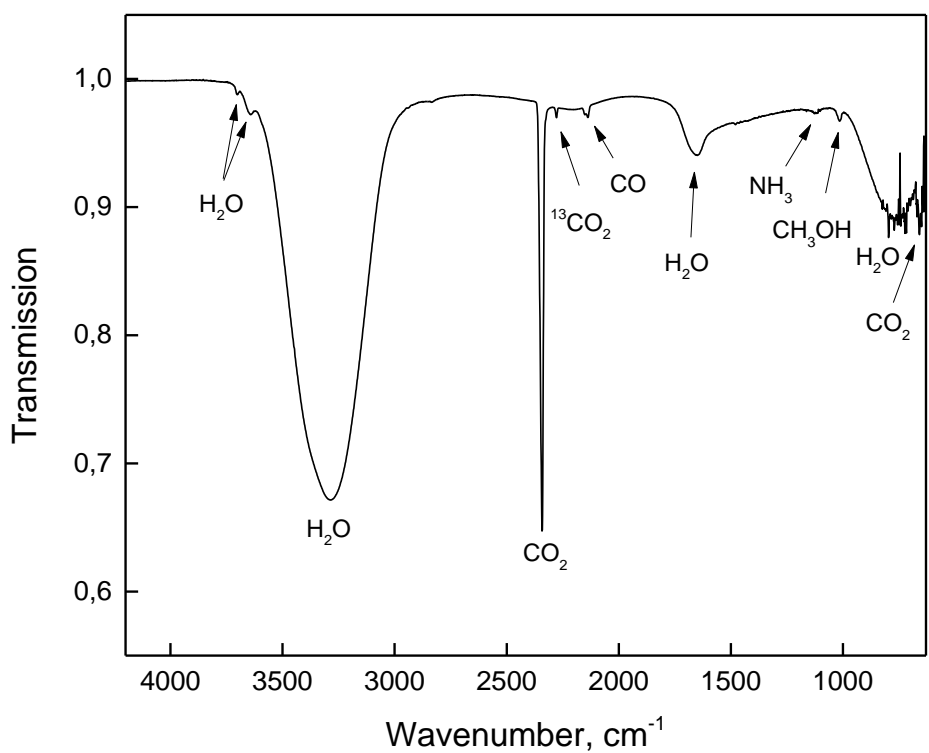

Figure 10. IR spectrum of a mixture of six ices (Exp3 in Table 2). The $O_2$ band is not visible.

Table 3. Frequencies and assignments of the vibrational bands observed in Fig.11.

| Frequency, cm$^{-1}$ | Assignment |
|---|---|
| 3700 | $H_2O$, dangling O–H and $CO_2$, combination mode |
| 3649 | $H_2O$, dangling O–H |
| 3285 | $H_2O$, stretching O–H |
| 2342 | $CO_2$, stretching C=O |
| 2279 | $^{13}CO_2$, stretching C=O |
| 2138 | CO, stretching C=O |
| 2200 | $H_2O$, combination mode* |
| 1660 | $H_2O$, bending |
| 1125 | $NH_3$, umbrella motion |
| 1013 | $CH_3OH$, stretching C–O |
| 760 | $H_2O$, librational motion |
| 658 | $CO_2$, bending |

*This band is a broad feature between 2500 and 2000 cm$^{-1}$.

After the depositions, TPD experiments were performed by linear ramping of the sample temperature with a rate of 2 K minute$^{-1}$ in the temperature range between 10 and 200 K. Infrared spectra were measured during the warming-up in the spectral range from 6000 to 600 cm$^{-1}$ with a resolution of 2 cm$^{-1}$. Mass spectra were taken during the warming-up in the mass range from 0 to 90 amu with a scanning time of one minute for one spectrum.

A very detailed comparison of the TPD behaviour of a five-component ice mixture to the TPD behaviour of its individual components was done recently (Martin-Domenech, et al. 2014). Here we compared five- and six-component ice mixtures, to understand if there is an influence of the additional $O_2$ in the mixture on the desorption behaviour of the composing molecules following the discussion in the introduction.

The MS TPD curves obtained during warming-up of the mixture of six ices containing 4% of $O_2$ are presented in Figure 11. Partial pressures are presented for six masses: 17 (OH and $NH_3$), 18 ($H_2O$), 28 (CO), 31 ($CH_3OH$), 32 ($O_2$ and $CH_3OH$), and 44 ($CO_2$).

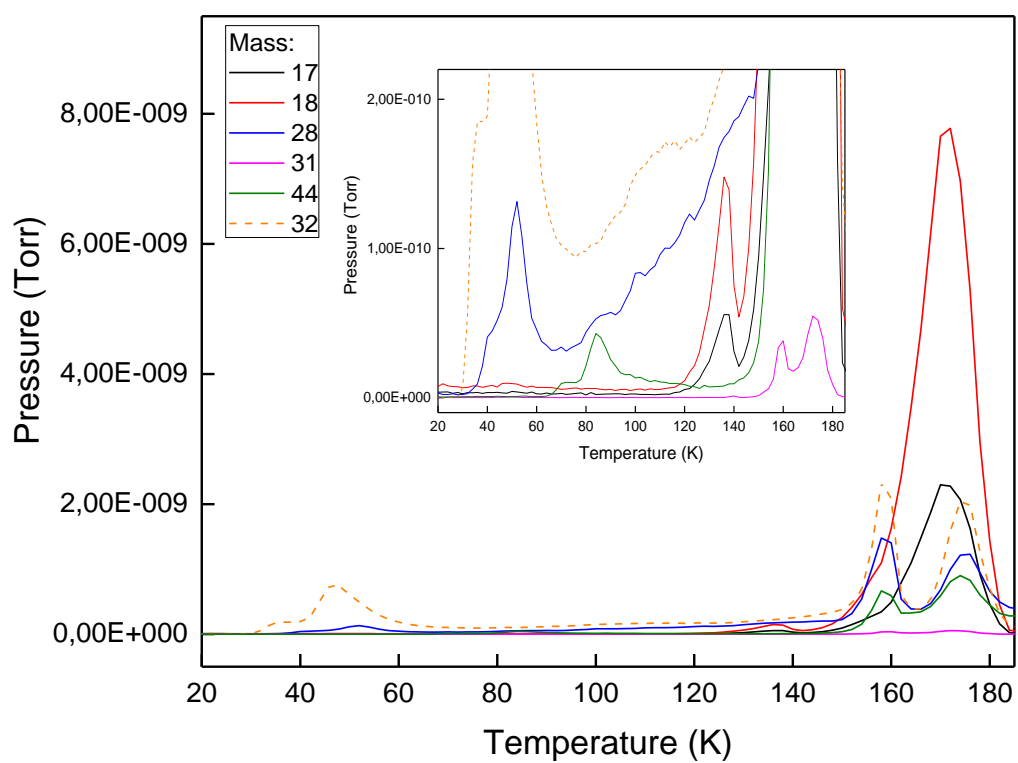

Figure 11. MS TPD curves obtained during warming-up of a mixture of six ices containing 4% of $O_2$ (Exp3 in Table 2). Inset: zoom-in.

Variations in the TPD peak positions for the ice mixtures containing 0, 1, and 4% of $O_2$ are in the limits of 1 K, which can be addressed to experimental uncertainties and measurement errors. However, a noticeable shift of the $O_2$ signal and additional TPD peaks are observed for the ice mixture containing 10% of $O_2$. TPD peak temperatures, corresponding molecules, and desorption types are listed in Table 4. All temperatures are in agreement with the literature data (Collings et al. 2004; Martin-Domenech, et al. 2014). TPD peaks at temperatures above the desorption temperatures corresponding to the pure ices are explained by trapping of species in the water-rich matrix. The $NH_3$ signal is indistinguishable from OH, which has, in addition, a much stronger signal.

Additional TPD peaks indicate that at high $O_2$ concentrations in cometary or interstellar ices, "heavy" ice molecules, such as $CO_2$ and $CH_3OH$, can be partly trapped in $O_2$ and release to the gas phase together with $O_2$ at temperatures much lower than expected. The amount of $O_2$ molecules desorbing independently relative to the amount of molecules desorbing due to their trapping in water matrix (volcano desorption and co-desorption) increases from ~10% for the 1% $O_2$ mixture to ~60% for the 10% $O_2$ mixture. The amounts of $CH_3OH$ and $CO_2$ molecules co-desorbing with $O_2$ relative to the $H_2O$-related desorption are about 2% and 1% correspondingly. For methanol, this gives ~$2\times10^{-3}$ $CH_3OH$ molecules per $O_2$ molecule, which is much more than the amount of methanol co-desorbing with CO (Ligterink et al. 2018). Thus, co-desorption of molecules with $O_2$ could partly explain the detection of $CH_3OH$ (Vastel et al. 2014; Potapov et al. 2016) and other complex organic molecules in cold dense molecular clouds, if we assume a relatively high amount of $O_2$ (similar to the comet 67P) in interstellar ices.

Table 4. TPD peak temperatures*, corresponding molecules and desorption types.

| TPD peak temperature, K (0, 1, 4% of $O_2$) | TPD peak temperature, K (10% $O_2$) | Molecule | Desorption type** |
|---|---|---|---|
| | 33 | $O_2$ | |
| | | $CH_3OH$ | co-desorption with $O_2$ |
| 36 | | $O_2$ | |
| 40 | 40 | CO | |
| | 41 | $O_2$ | |
| | | $CO_2$ | co-desorption with $O_2$ |
| | | $CH_3OH$ | co-desorption with $O_2$ |
| 47 | | $O_2$ | |
| 51 | 51 | CO | |
| 70 | 70 | $CO_2$ | |
| 85 | 85 | $CO_2$ | |
| 136 | 136 | $H_2O$ | co-desorption with $CH_3OH$ |
| 138 | 138 | $CH_3OH$ | |
| 157 | 157 | CO, $CO_2$, $O_2$, $CH_3OH$ | volcano desorption of molecules trapped in water ice matrix due to the transformation of the water ice structure from amorphous to crystalline |
| 171 | 171 | $H_2O$ | |
| 173 | 173 | CO, $CO_2$, $O_2$, $CH_3OH$ | co-desorption with $H_2O$ |

*Variations in the TPD peak positions for different ice mixtures are in the limits of 1 K, which can be addressed to experimental uncertainties and measurement errors.

**From our point of view, in such complicated mixtures with many interactions between molecules in the water ice matrix, which can have a high porosity and consequently a large surface, there is no possibility to distinguish between monolayer and multilayer desorption.

**Conclusions**

A new experimental set-up has been developed in our laboratory and described in this paper. INSIDE (INterStellar Ice Dust Experiment) allows the studying physical and chemical processes on the surface of bare or ice-covered cosmic dust grain analogues.

Photodesorption of water ice molecules deposited onto the surface of amorphous silicate and carbon grains was measured for the first time. The photodesorption yield for 66 ML of $H_2O$ ice on $MgFeSiO_4$ grains is in agreement with the literature data indicating that water molecules are weakly bound on the surface of silicates. Much lower yield for 3 ML of $H_2O$ ice on $MgFeSiO_4$ grains is in agreement with our previous study demonstrated that ice molecules desorb from a surface of grain clusters that is much larger than the nominal surface of the substrate. The photodesorption yield for $H_2O$ on carbon grains is noticeably higher as compared to silicate grains. This result could be explained by the hydrophobic surface properties of the carbon grains and a stronger absorption of UV photons by carbon increasing the surface temperature.

Temperature programmed desorption of a six-component ice ($H_2O$, CO, $CO_2$, $CH_3OH$, $NH_3$, and $O_2$) was measured for the first time. No changes in the TPD peaks positions of the ice components due to an addition of up to 10% of $O_2$ to the ice mixture have been observed. Co-desorption of $CO_2$ and $CH_3OH$ with $O_2$ in the ice containing 10% of $O_2$ indicates that at high $O_2$ concentrations in cometary or interstellar ices "heavy" ice molecules can be partly trapped in $O_2$ and release to the gas phase much earlier than expected. This effect could explain astronomical detections of complex organic molecules in cold dense interstellar clouds. The detection of a "new" molecule formed by thermal reactions in complex ice mixtures is an interesting result, which needs further investigations.

**Acknowledgments**

We would like to thank the anonymous referee for questions, suggestions, and corrections that helped to improve the manuscript. This work was partly supported by the Research Unit FOR 2285 "Debris Disks in Planetary Systems" of the Deutsche Forschungsgemeinschaft (grant JA 2107/3-1).